\documentclass[secnumarabic,amssymb, nobibnotes, aps, prd]{revtex4-2}
\usepackage{graphicx}
\usepackage{amsmath}
\usepackage{subcaption}
\usepackage{xcolor}
\usepackage{multirow}

\begin{document}
	\title{Restricted Phase Space Thermodynamics of Charged Static and Charged Rotating Black Holes in $f(R)$ Gravity}
	\author{Amijit Bhattacharjee$^1$}
	
	\email{$rs_amijitbhattacharjee@dibru.ac.in$}
	\author{Prabwal Phukon$^{1,2}$}
	\email{prabwal@dibru.ac.in}	
	\affiliation{$1.$Department of Physics, Dibrugarh University, Dibrugarh, Assam,786004.\\$2.$Theoretical Physics Division, Centre for Atmospheric Studies, Dibrugarh University, Dibrugarh, Assam,786004.}
	\begin{abstract}

The thermodynamics of black holes provides a profound link between gravity, quantum theory and statistical mechanics. It serves as a useful tool for testing theories beyond Einstein's gravity. In this work of ours, we investigate the newly found restricted phase space thermodynamics (RPST) of charged static and charged rotating black holes in $f(R)$ gravity. Unlike the extended phase space (EPST) approach, RPST keeps the cosmological constant fixed and introduces the central charge $C$ along with its conjugate chemical potential $\mu$, thereby allowing the black hole mass to be consistently interpreted as internal energy. Within this framework, we derive the relevant thermodynamic quantities and analyse the temperature-entropy $(T-S)$ and Helmholtz free energy-temperature $(F-T)$ behaviours. Our results reveal characteristic features of first-order phase transitions through non-monotonic $T-S$ curves along with the swallow-tail structures in $F-T$ plots, while second-order transitions appear at critical points. To further validate these findings, we employ the formalism of geometrothermodynamics (GTD), which provides a Legendre-invariant geometric description of thermodynamic geometry. We demonstrate that the curvature singularities of the GTD scalar curvature coincides exactly with that of the divergences in the specific heat capacity curves, thereby establishing a geometric correspondence for phase transitions. This study facilitates the first systematic exploration of RPST within $f(R)$ gravity and highlights the universality of RPST in capturing black hole criticality in modified gravity theories.

\end{abstract}

	\maketitle
	\section{Introduction}

The thermodynamics of black holes has become a cornerstone in our understanding of the interplay between gravitation, quantum theory, and statistical mechanics. Since the pioneering works of Bekenstein and Hawking \cite{Bekenstein:1973ur,Hawking:1974rv}, it has been firmly established that black holes behave as thermodynamic systems endowed with temperature and entropy, obeying laws that closely parallel the conventional laws of thermodynamics. These insights not only provide an avenue for probing the microscopic structure of spacetime, but also establish black holes as ideal laboratories for testing gravitational theories beyond general relativity.\\

Much of the existing literature has been developed within the framework of extended phase space thermodynamics (EPST), wherein the cosmological constant is promoted to a thermodynamic variable identified as pressure, and the black hole mass is reinterpreted as enthalpy \cite{Kubiznak:2012wp}. This approach has revealed striking analogies between black hole systems and classical fluids, most notably the Van der Waals-like phase transitions. Nevertheless, an alternative perspective, known as restricted phase space thermodynamics (RPST), has been put forward \cite{Cong:2021fnf}, in which the black hole mass retains its role as internal energy and the cosmological constant is kept fixed. RPST provides a conceptually clean framework for analyzing black hole phase transitions and has been successfully applied to various asymptotically AdS and non-AdS backgrounds \cite{Wang:2022pkj,Huang:2024jzz,Kong:2023rpst}. This motivates further exploration of RPST within the broader class of modified gravity theories.\\

Unlike EPST where the cosmological constant is considered to be dynamical, RPST \cite{Cong2021, Visser2022} on the other hand introduces a new pair of thermodynamic conjugates: the central charge $ C $, which quantifies the number of microscopic degrees of freedom in the dual conformal field theory, along with its conjugate chemical potential $ \mu $. In this formulation, the black hole mass is again reinterpreted as the internal energy, and the thermodynamic structure reflects that of the holographic principles more directly. The thermodynamic variables $ P $ and $ V $ are constrained in RPST, which leads to a simplified first law of black hole thermodynamics:
\begin{equation}
\label{eq1}
dM = TdS + \Omega dJ + \tilde{\Phi} d\tilde{Q} + \mu dC,
\end{equation}

where $l$, the AdS radius is kept fixed which in turn eliminates any volume work thus simplifying the analysis of black hole thermodynamic processes.

Despite the rapid development of restricted phase space thermodynamics in the context of Einstein gravity, its application to modified gravity theories has remained largely unexplored. Most of the existing studies on RPST have been restricted to asymptotically AdS black holes within general relativity \cite{Cong:2021fnf,Wang:2022pkj,Huang:2024jzz,Kong:2023rpst}, leaving open the question of whether the structural features of RPST extend to broader gravitational frameworks. In particular, theories such as $f(R)$ gravity, which naturally arise as extensions of the Einstein--Hilbert action and are motivated by cosmological observations and quantum corrections, provide an ideal ground to test the universality of RPST. To the best of our knowledge, the present work constitutes the first systematic attempt to investigate the restricted phase space thermodynamics of charged static and charged rotating black holes in $f(R)$ gravity. Our analysis thus not only extends the applicability of RPST to modified gravity but also uncovers new insights into the phase structure and criticality of black holes beyond Einstein’s theory.\\

f(R) gravity is a particularly appealing modification of general relativity, where the Einstein–Hilbert Lagrangian is generalized to an arbitrary function of the Ricci scalar $R$ \cite{Sotiriou:2008rp,DeFelice:2010aj}. Such theories naturally emerge as effective descriptions in quantum gravity scenarios and string-inspired models, and they have been extensively studied as viable candidates to explain cosmic acceleration without invoking dark energy. Black hole solutions in f(R) gravity provide valuable insights into how modifications to Einstein’s theory affect horizon structure, thermodynamic quantities, and critical phenomena \cite{Cognola:2005de,Myung:2011we}. Therefore, examining black hole thermodynamics in f(R) gravity within the RPST framework constitutes an important step in understanding the universality and robustness of these approaches.\\

In this work, we investigate the restricted phase space thermodynamics of charged static and charged rotating black holes in f(R) gravity. Our analysis demonstrates rich phase structures in both cases. In particular, the temperature–entropy ($T$–$S$) diagrams exhibit nonmonotonic behavior, signaling the occurrence of first-order phase transitions. Consistently, the free energy–temperature ($F$–$T$) plots display the characteristic swallow-tail profile, a hallmark of first-order phase transitions in thermodynamic systems, including black holes. These results confirm that the RPST framework is capable of capturing criticality in black holes beyond the scope of Einstein gravity.\\

To further substantiate our analysis, we employ the geometric approach of geometrothermodynamics (GTD) \cite{Quevedo2007,Quevedo2008}. The GTD metric is a Legendre invariant metric and therefore would not depend on any specific choice of thermodynamic potential. The phase transitions obtained from the specific heat capacity of the black hole are properly contained in the scalar curvature of the GTD metric, such that a curvature singularity in the GTD scalar `$R_{GTD}$'  would imply the occurrence of a phase transition. The general form of the type II GTD metric is given by \cite{{Soroushfar2016}} :
\begin{equation}
\label{eq2}
g = \left(E^{c} \frac{\partial{\varphi}}{\partial{E^{c}}} \right)
\left(\eta_{ab} \delta^{bc} \frac{\partial^2 \varphi}{\partial E^{c} \partial E^{d}} dE^{a} dE^{d}\right)
\end{equation}     
where `$\varphi$' is the thermodynamic potential and `$E^a$' is an extensive thermodynamic variable with $a=1,2,3....$..

 The thermodynamic scalar curvature, computed within GTD, is shown to diverge exactly at the points where the specific heat diverges, thus establishing a precise correspondence between geometric and thermodynamic signals of phase transitions. This similarity strengthens the interpretation of GTD as a powerful diagnostic tool for black hole criticality in modified gravity theories.\\

The remainder of the paper is organized as follows. In Section~\ref{sec:background}, we provide a brief overview of charged static black hole solutions in f(R) gravity and outline the RPST formalism along with the $T$–$S$ and $F$–$T$ behaviours. Section~\ref{sec:thermodynamics} presents the thermodynamic analysis of charged rotating black holes, including the $T$–$S$ and $F$–$T$ criticality plots. In Section~\ref{sec:GTD}, we analyze the GTD geometry of both the charged static and charged rotating black holes in f(R) gravity and its relation to phase transitions. Finally, Section~\ref{sec:conclusion} summarizes our findings and discusses their broader implications.  

\section{Charged Static Black Hole in $f(R)$ Gravity}
\label{sec:background}

The charged static black hole solution in $f(R)$ gravity that we have considered here originates from the following action \cite{{Soroushfar2016}}
	$$S=\frac{1}{16 \pi G} \int d^4x \sqrt{-g} \left(R+f(R)-F_{\mu \nu} F^{\mu \nu}\right)$$
	Varying the action with respect to the metric gives :
	$$ R_{\mu \nu}(1+f'(R)) - \frac{1}{2} \left(R+f(R)\right) g_{\mu \nu}+\left(g_{\mu \nu} \nabla^2-\nabla _\mu \nabla_\nu \right) f'(R)=2T_{\mu \nu}$$
	where $f'(R)=\frac{df(R)}{dR}$ and $T_{\mu \nu}$ is the stress-energy tensor of the electromagnetic field.\\
	
	The trace of the above equation at $R=R_0$ results in :
	$$R_0(1+f'(R_0))-2((R_0+f(R_0))=0$$
	which eventually gives the constant curvature scalar as:
	$$R_0=\frac{2 f(R_0)}{f'(R_0)-1}$$
	Finally, the metric of the spherically symmetric space time is obtained as follows:
	\begin{equation}
		ds^2=P(r)dt^2-\frac{1}{P(r)} dr^2-r^2(d\theta^2+sin^2\theta d\phi^2)
		\label{eq3}
	\end{equation}
	Where,
	\begin{equation}
		P(r)=1-\frac{2 G M}{r}+\frac{Q^2}{\left(1+f'(R_0)\right) r^2}-\frac{R_0 r^2}{12}
	\end{equation}
	for the details about the metric see \cite{Moon:2011}.\\
	
	By putting  $\left(1+f'(R_0)\right)=b$ in the above equation:
	\begin{equation}
		P(r)=1-\frac{2 G  M}{r_+}+\frac{G Q^2}{b r_+^2}-\frac{R_0 r_+^2}{12}
		\label{eq4}	
	\end{equation}
	where $R_0=4 \Lambda$=$-\frac{12}{l^2}$ is the constant curvature. For charged static black hole, from the equation \ref{eq4},mass $M$ in canonical ensemble is obtained as \cite{{Soroushfar2016}} :
	\begin{equation}
		M = \frac{  G b S^{2} + l^{2} \pi \left( \pi Q^{2} + b  S \right)}{2 b l^{2} \pi^{3/2} \sqrt{\left( G S \right)}}
\label{eq5}
	\end{equation}
where, $S=\frac{\pi r_+^2}{G}$ and now in the Restricted phase space formalism, the following replacements are to be made:
	\begin{equation}
\tilde{Q}= \frac{Q l}{ \sqrt{G}}~~~~~~~and~~~~~~~~ C= \frac{l^2}{G}
\end{equation}
and following which equation.\ref{eq5} can be rewritten as:
\begin{equation}
M = \frac{   b S^{2} +  \pi \tilde{Q}^{2} + \pi b C  S }{2 b l \pi^{3/2} \sqrt{\left( C S \right)}}
\label{eq6}
\end{equation}
If $S$, $\tilde{Q}$ and $C$ are rescaled as $S \to \gamma S$, $\tilde{Q} \to \gamma \tilde{Q}$ and $C \to \gamma C$  then eqtn.(\ref{eq6}) implies  $M \to \gamma M$  which proves the first order homogeneity of $M$. Now, owing to the expression for mass in eqtn. (\ref{eq6}) for the charged static black hole in the restricted phase space formalism, the following thermodynamic quantities can be computed as:

\begin{equation}
T = \frac{3 b \sqrt{C}\, S^{5/2} - \pi^{2} \tilde{Q}^{2} \sqrt{C S} + b C \pi S \sqrt{C S}}{4 b C l \pi^{3/2} S^{2}}
\label{eq7}
\end{equation}

\begin{equation}
SH = \frac{2 S \left(-\sqrt{C}\, \pi^{2} \tilde{Q}^{2} \sqrt{C S} 
   + b S \sqrt{C S} \left(C^{3/2} \pi + 3 \sqrt{S} \sqrt{C S}\right)\right)}
{3 \sqrt{C}\, \pi^{2} \tilde{Q}^{2} \sqrt{C S} 
   + b S \sqrt{C S} \left(-C^{3/2} \pi + 3 \sqrt{S} \sqrt{C S}\right)}
   \label{eq8}
\end{equation}

\begin{equation}
\mu = \frac{S \left( \pi b C S - \pi^2 \tilde{Q}^2 - b S^2 \right)}
{4 l b \, (\pi C S)^{3/2}}
\label{eq9}
\end{equation}

The temperature, specific heat capacity and the chemical potential  can be respectively seen from the above equations i.e. eqtns.(\ref{eq7})---(\ref{eq9}) to distinctly show the zeroth order homogeneity for all the above mentioned quantities.

\subsection{T-S and F-T plots of the iso-e-charge process for charged static black hole} 

In this section, we will look into the various allowed thermodynamic processes of the black hole namely: the $T-S$ and the $F-T$ processes. In order to obtain the critical points, we solve the following pair of equations which are given by :
\begin{equation}
\frac{d T}{d S}=0,~~~~~~~~~~~~\frac{d^2 T}{d S^2}=0
\end{equation}
Solving the above mentioned equations we obtain the critical value of $S$ and $\tilde{Q}$ which are given as:
\begin{equation}
S_C=\frac{C \pi}{6} ~~~~~~~~~~\tilde{Q}_C= \frac{\sqrt{b} C}{6} 
\label{eq10}
\end{equation}
By substituting the critical values from eqtn.\ref{eq9} in the expression for $T$ in eqtn.\ref{eq7}, the critical temperature,
 $ T_C$ is found to be $ T_C=0.259899$, and with
\begin{equation}
t=\frac{T}{T_C},~~~~s=\frac{S}{S_C},~~~~~q =\frac{\tilde{Q}}{\sqrt{b}~\tilde{Q}_C}
\end{equation}

and by putting $b=2$ using which the relative temperature, $t$ can be written as:
\begin{equation}
t = \frac{3 s^{\frac{5}{2}} + \sqrt{s}(6s - q^2)}{8 s^2}
\end{equation}

and the rescaled Helmholtz free energy $f=F/F_C$ is obtained to be
\begin{equation}
f = \frac{ 3C (q^2 + 2s) - C s^2}{\sqrt{s}(8C) }
\end{equation}
It is important to mention that for $b > 1$ we enter in the realm of f(R) modified gravity away from the Einstein's theory.
\begin{figure}[ht]
\begin{center}
\includegraphics[width=.52\textwidth]{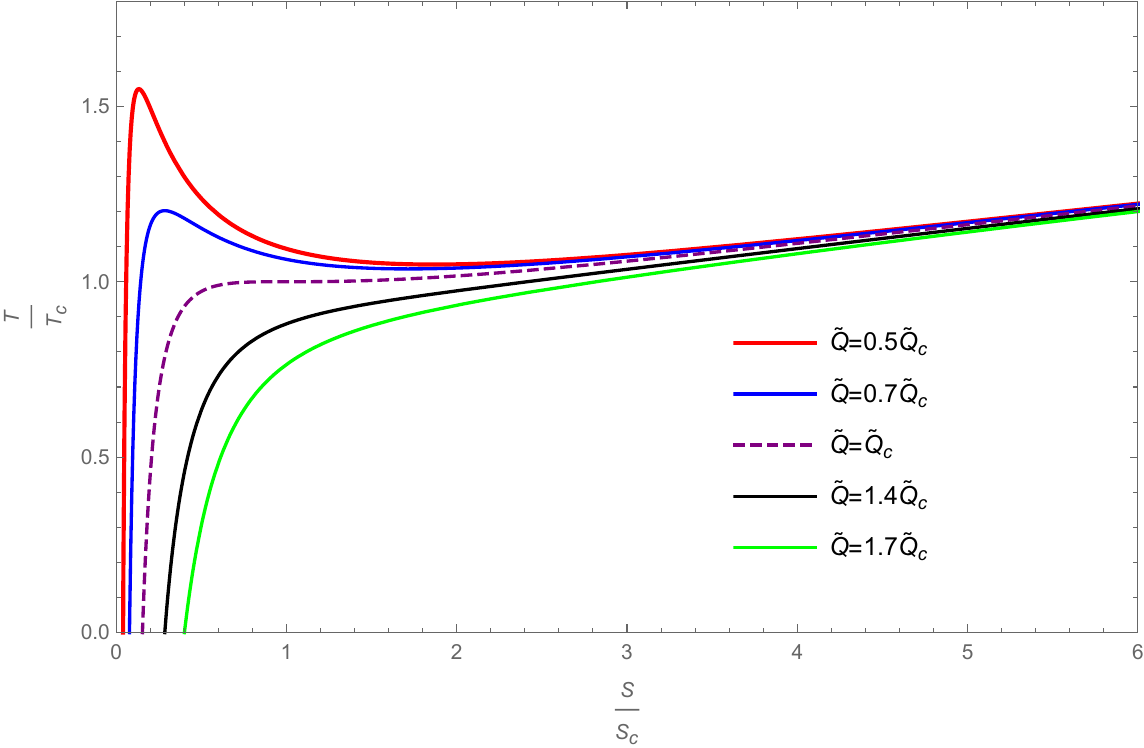}\vspace{3pt}
\includegraphics[width=.48\textwidth]{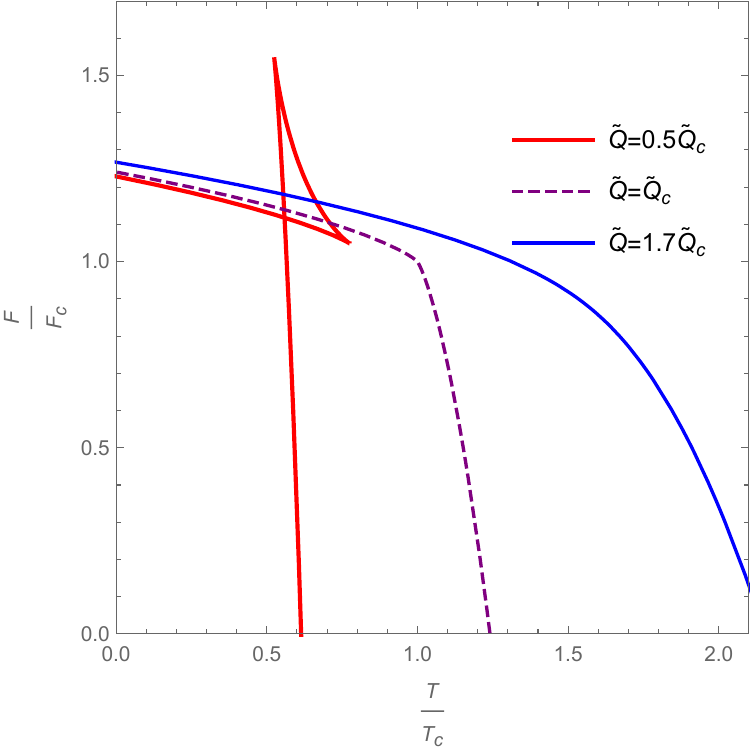}
\caption{$T-S$ and $F-T$ curves for charged static black hole in the iso-$e$-charge processes for $C=53$ and $b=2$.}\label{fig1}
\end{center}
\end{figure}
 FIG. \ref{fig1} here illustrates both the $T-S$ and $F-T$ curves for the iso-e-charge process for the charged static black hole in f(R) gravity. The plots quite distinctly show that below the critical value of $\tilde{Q}_C$ both the $T-S$ and $F-T$ curves show non-trivial behaviour where, the $T-S$ curve exhibits non-monotonic behaviour below $\tilde{Q}_C$ whereas the $F-T$ curve displays a swallowtail behaviour below the critical value of $\tilde{Q}_C$. Such behaviour is an indication for the possible Van der Waals-like first-order phase equilibrium in the iso-e-charge processes when $0 < \tilde{Q} < \tilde{Q}_C$. However at the critical point $\tilde{Q} = \tilde{Q}_C$ we see that the second order phase transition is observed as represented by the dashed solid curve in both the figures.\\

\begin{figure}[ht]
\begin{center}
\includegraphics[width=.48\textwidth]{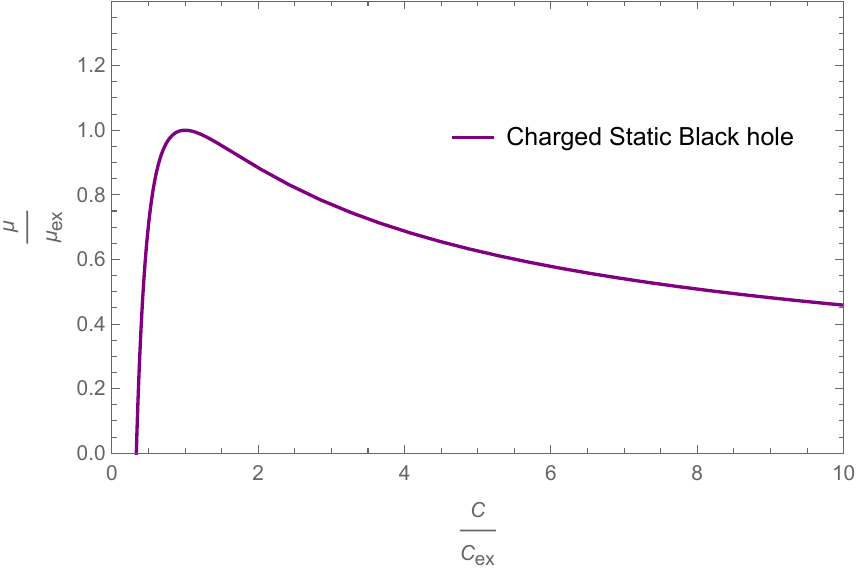}
\caption{$\mu-C$ curves for charged static black hole in the iso-$e$-charge processes for $C=53$ and $b=2$.}\label{fig2}
\end{center}
\end{figure}

We then analyze the $\mu-C$ process in detail for the iso-e-charge process. We numerically compute the extrememum value of $\mu$ i.e. $\mu_{ex}$  using eqtn.(\ref{eq9}), keeping $\tilde{Q}$,$b$, $C$ and $S$ fixed. We then draw a plot between $\frac{\mu}{\mu_{ex}}$ versus $\frac{C}{C_{ex}}$ and the corresponding plot is shown as  Fig.\ref{fig2}.

\section{Charged Rotating Black Hole in $f(R)$ Gravity}
\label{sec:thermodynamics}
In this section, we study the thermodynamics of a charged rotating  black hole solution in $f(R)$ gravity and the action that facilitates the required black hole solution is given as:
	\begin{equation}
		S=\frac{1}{16 \pi}\left(\int d^D x \sqrt{|g|} (R+f(R))-\int d^4 x  \sqrt{|g|} [F_{\mu \nu}F^{\mu \nu}]\right)
	\end{equation}
	where the first part is gravitational whereas the second part represents the four dimensional Maxwell term. $R$ is the scalar curvature and $R+f(R)$ is a function of scalar curvature.The field equations in the metric formalism are \cite{Larraaga:2012}
	\begin{equation}
		R_{\mu \nu}(1+f'(R))-\frac{1}{2}(R+f(R)) g_{\mu \nu} +(g_{\mu \nu} \nabla^2 -\nabla_\mu \nabla_\nu)f'(R)=2 T_{\mu \nu}
		\label{eq11}
	\end{equation}
	where, $\nabla$ is the covariant derivative,$R_{\mu \nu}$ is the Ricci tensor, and $T_{\mu \nu}$  is the stress-energy tensor of the electromagnetic field given by:
	$$T_{\mu \nu}=F_{\mu \rho} F^\rho_\nu-\frac{g_{\mu \nu}}{4} F_{\rho \sigma}F^{\rho \sigma}$$
	The trace of equation \ref{eq11} gives the expression for  $R=R_0$ as:
	\begin{equation}
		R_{0}=\frac{2 f(R_0)}{f'(R_0)-1}
	\end{equation}
	The axisymmetric ansatz, utilizing Boyer-Lindquist type coordinates $(t, r, \theta, \varphi)$, derived from the Kerr-Newman-Ads black hole solution, as shown in \cite{Larraaga:2012} is:
	\begin{equation}
		ds^2=-\frac{\Delta_r}{\rho^2}\left [ dt-\frac{a sin^2\theta d\varphi}{\Xi} \right ]^2+\frac{\rho^2}{\Delta_r} dr^2 +\frac{\rho^2}{\Delta_\theta} d\theta^2+\frac{\Delta_\theta sin^2 \theta} {\rho^2}\left[a dt-\frac{r^2+a^2}{\Xi} d\varphi\right]^2
		\label{eq12}
	\end{equation}
	where 
	$$\Delta_{r}=(r^2+a^2)\left(1+\frac{R_0}{12} r^2 \right)-2m Gr+\frac{Q^2}{(1+f'(R_0))},$$
	$$\Xi=1-\frac{R_0}{12}a^2 \hspace{0.5cm},\hspace{0.5cm} \rho^2=r^2+a^2cos^2\theta,$$
	$$\Delta_\theta=1-\frac{R_0}{12}a^2 cos^2\theta$$
	in which $R_0=-4 \Lambda$, Q is the electric charge and $a$ is the angular momentum per mass of the black hole. By setting $dr=dt=0$ in equation \ref{eq12}, we can calculate the area of the two-dimensional horizon, which eventually gives the expression for area as\cite{{Soroushfar2016}}:
	\begin{equation}
		S=\frac{\pi(r_+^2+a^2)}{G(1-\frac{R_0}{12}a^2)}
	\end{equation}
	where $r_+$ is the radius of the horizon.\\
	The expressions for total mass and the angular momentum are \cite{Larraaga:2012} :
	\begin{equation}
		M=\frac{m}{\Xi^2}
	\end{equation}
	and
	\begin{equation}
		J=\frac{a m}{\Xi^2}
	\end{equation}
	From which the generalized Smarr formula of the rotating charged black hole is obtained as\cite{Larraaga:2012}:
	\begin{equation}
		M^2=\frac{S}{4 \pi G }-\frac{J^2 R_0}{12}+\frac{\pi  \left(4 J^2+\frac{Q^4}{b^2}\right)}{4 S G}-\frac{R_0 S \left(\frac{Q^2}{b}-\frac{R_0 G S^2}{24 \pi ^2}+\frac{S}{\pi }\right)}{24 \pi }+\frac{Q^2}{2b G}
		\label{smr}
		\end{equation}
	
	In the fixed $(q, J)$ ensemble, we keep $q$ and $J$ as fixed parameters. The mass expression, given by equation \ref{smr}, remains unchanged within this ensemble,which is :
	\begin{equation}
		M=\sqrt{\frac{S}{4 \pi G }-\frac{J^2 R_0}{12}+\frac{\pi  \left(4 J^2+\frac{Q^4}{b^2}\right)}{4 S G}-\frac{R_0 S \left(\frac{Q^2}{b}-\frac{R_0 G S^2}{24 \pi ^2}+\frac{S}{\pi }\right)}{24 \pi }+\frac{Q^2}{2b G}}
	\label{eq13}
	\end{equation}
And with the following replacements the mass for the charged rotating black hole could be written in restricted phase space formalism:
	\begin{equation}
\tilde{Q}= \frac{Q l}{ \sqrt{G}}~~~~~~~and~~~~~~~~ C= \frac{l^2}{G}
\label{eq14}
\end{equation}

Combing equation. \ref{eq13} with equation. \ref{eq14} and after putting $R_0=4 \Lambda$=$-\frac{12}{l^2}$ we get:
\begin{equation}
		M=\sqrt{\frac{S C}{4 \pi l^2 }+ J^2 +\frac{\pi C  \left(4 J^2+\frac{\tilde{Q}^4}{C^2 b^2}\right)}{4 S l^2}+ \frac{ S \left(\frac{\tilde{Q}^2}{C b}+\frac{  S^2}{2 C \pi ^2}+\frac{S}{\pi }\right)}{2 \pi }+\frac{\tilde{Q}^2}{2b l^2}}
	\label{eq15}
	\end{equation}

If $S$, $\tilde{Q}$, $J$ and $C$ are rescaled as $S \to \gamma S$, $\tilde{Q} \to \gamma \tilde{Q}$, $J \to \gamma J$ and $C \to \gamma C$  then eqtn.(\ref{eq15}) implies  $M \to \gamma M$  which proves the first order homogeneity of $M$. Now, owing to the expression for mass in eqtn. (\ref{eq15}) for the charged rotating black hole in the restricted phase space formalism, the following thermodynamic quantities can be computed as:

\begin{equation}
T = \frac{C \left( 3 S^{4} + 4 \pi S^{3} C + \pi^{2} S^{2} \left( 2 \tilde{Q}'^{2} + C^{2} \right) - \pi^{4} \left( \tilde{Q}'^{4} + 4 J^{2} C^{2} \right) \right)}{4 l \pi^{3/2} \left( S C \right)^{3/2} \sqrt{ 4 J^{2} \pi^{3} C \left( S + \pi C \right) + \left( \pi^{2} \tilde{Q}'^{2} + S^{2} + \pi S C \right)^{2} }}
\label{eq16}
\end{equation}

\begin{equation}
\mu = \frac{S \left( - S^{4} + \pi^{2} S^{2} \left( -2 (\tilde{Q}')^{2} + C^{2} \right) - \pi^{4} \left( (\\tilde{Q}')^{4} - 4 J^{2} C^{2} \right) \right)}
{4 l \pi^{3/2} \left( S C \right)^{3/2} \sqrt{ 4 J^{2} \pi^{3} C (S + \pi C) + \left( \pi^{2} (\tilde{Q}')^{2} + S^{2} + \pi S C \right)^{2} }}
\label{eq17}
\end{equation}

\begin{equation}
SH = \frac{2 S \left( 3 S^{4} + 4 \pi S^{3} C + \pi^{2} S^{2} (2 (\tilde{Q}')^{2} + C^{2}) - \pi^{4} ((\tilde{Q}')^{4} + 4 J^{2} C^{2}) \right) 
\mathcal{A}}
{3 S^{8} + 24 \pi^{3} (2 J^{2} + (\tilde{Q}')^{2}) S^{5} C + 8 \pi S^{7} C + 6 \pi^{2} S^{6} (2 (\tilde{Q}')^{2} + C^{2}) + 8 \pi^{7} (2 J^{2} + (\tilde{Q}')^{2}) S C ((\tilde{Q}')^{4} + 4 J^{2} C^{2}) + \mathcal{B}}
\label{eq18}
\end{equation}

where, $\mathcal{A}=\left( S^{4} + 2 \pi^{3} (2 J^{2} + (\tilde{Q}')^{2}) S C + 2 \pi S^{3} C + \pi^{2} S^{2} (2 (\tilde{Q}')^{2} + C^{2}) + \pi^{4} ((\tilde{Q}')^{4} + 4 J^{2} C^{2}) \right)$ and,\\

$\mathcal{B}=24 \pi^{5} S^{3} C ((\tilde{Q}')^{4} + 4 J^{2} C^{2}) + 6 \pi^{6} S^{2} (2 (\tilde{Q}')^{2} + C^{2})((\tilde{Q}')^{4} + 4 J^{2} C^{2}) + 3 \pi^{8} ((\tilde{Q}')^{4} + 4 J^{2} C^{2})^{2} + \pi^{4} S^{4} (18 (\tilde{Q}')^{4} + 120 J^{2} C^{2} + 12 (\tilde{Q}')^{2} C^{2} - C^{4})$\\

Here,$\tilde{Q}'= \frac{\tilde{Q}}{\sqrt{b}}$ and the temperature, chemical potential and the specific heat capacity can be  respectively seen from the above equations i.e. eqtns.(\ref{eq16})---(\ref{eq18}) to distinctly show the zeroth order homogeneity for all the above mentioned quantities.
\subsection{T-S and F-T plots of the iso-e-charge process for charged rotating black hole} 

In this section, we will look into the various allowed thermodynamic processes of the black hole namely: the $T-S$ and the $F-T$ processes. In order to obtain the critical points, we solve the following pair of equations which are given by :
\begin{equation}
\frac{d T}{d S}=0,~~~~~~~~~~~~\frac{d^2 T}{d S^2}=0
\label{eq19}
\end{equation}
Using equation.\ref{eq14} in equation.\ref{eq17} by differentiating and double differentiating T respectively, we get the following equations:
\begin{align}
&16 J^4 C^3 \pi^7 (3C\pi + 4S) 
+ (\pi^2 \tilde{Q}'^2 + C\pi S + S^2)^3 (3\pi^2 \tilde{Q}'^2 - C\pi S + 3S^2)\nonumber \\
&\quad  + 8 J^2 C \pi^3 (3C \pi^5 \tilde{Q}'^4 + 2\pi^4 \tilde{Q}'^2 (2C^2 + \tilde{Q}'^2) S \nonumber \\
&\quad + 3 C \pi^3 (C^2 + 2\tilde{Q}'^2) S^2 + 12C^2 \pi^2 S^3 + 15C \pi S^4 + 6 S^5) = 0,
\label{20}
\end{align}

\begin{align}
&-(5\pi^2 \tilde{Q}'^2 - C\pi S + S^2)(\pi^2 \tilde{Q}'^2 + C\pi S + S^2)^5 
- 64 J^6 C^4 \pi^{10}(5C^2 \pi^2 + 12C\pi S + 8S^2) \nonumber \\
&\quad - 16 J^4 C^2 \pi^6 (15C^2 \pi^6 \tilde{Q}'^4 + 24C \pi^5 \tilde{Q}'^2 (C^2 + \tilde{Q}'^2) S 
+ \pi^4 (13C^4 + 58C^2 \tilde{Q}'^2 + 8\tilde{Q}'^4) S^2 \nonumber \\
&\quad + 40 C \pi^3 (C^2 + \tilde{Q}'^2) S^3 + 35C^2 \pi^2 S^4 - 8S^6) 
- 4 J^2 C \pi^3 (\pi^2 \tilde{Q}'^2 + C\pi S + S^2)(15C \pi^7 \tilde{Q}'^6 \nonumber \\
&\quad + 3\pi^6 \tilde{Q}'^4 (11C^2 + 4\tilde{Q}'^2) S 
+ C \pi^5 \tilde{Q}'^2 (25C^2 + 57\tilde{Q}'^2) S^2 
+ \pi^4 (15C^4 + 56C^2 \tilde{Q}'^2 + 28\tilde{Q}'^4) S^3 \nonumber \\
&\quad + 5C \pi^3 (13C^2 + \tilde{Q}'^2) S^4 + 7\pi^2 (15C^2 - 4\tilde{Q}'^2) S^5 
+ 75C \pi S^6 + 20 S^7 ) = 0.
\label{21}
\end{align}

Here,$\tilde{Q}'= \frac{\tilde{Q}}{\sqrt{b}}$ and unlike the previous charged static case which consisted only of one characteristic parameter i.e. the scaled electric charge,$\tilde{Q}$. We here in the charged rotating black hole in f(R) gravity encounter two parameters instead. These parameters are the rescaled charge $\tilde{Q}$ and angular momentum $J$ which makes the analytical computation of the critical points impossible for the charged rotating black hole and therefore we move on to a numerical method for computing the critical parameters for the black hole system. We here in the charged rotating black hole along with entropy,S and central charge,C have four thermodynamic variables (S, $\tilde{Q}$, J, C) and only two constraint equations. We therefore need to reduce the number of variables to two instead of four so as to facilitate the numerical analysis. For this, we apply the method ued in the paper\cite{Tripathy:2024} by which we define a dimensionless parameter say $\sigma$ as the ratio of angular momentum,J and the rescaled charge,$\tilde{Q}$:

\begin{equation}
\sigma= \frac{J}{\tilde{Q}}
\end{equation}
And therefore instead of (J, $\tilde{Q}$), we now write the critical points in terms of ($\sigma, \tilde{Q}$). And through dimensional analysis we write:
\begin{equation}
S_c=f(\sigma)\tilde{Q},~~~~~~~~~~~~C=g(\sigma)\tilde{Q}
\label{eq22}
\end{equation}
After this the equations \ref{20} and \ref{21} reduce to the form comprising only of $f(\sigma)$ and $g(\sigma)$ and we solve them numerically and successfully find a fitting form allowing us to express both $f(\sigma)$ and $g(\sigma)$ as functions of $\sigma$. The form of $f(\sigma)$ and $g(\sigma)$ are found to be:
\begin{align}
f(\sigma) = 
\frac{
-1.688576109542579710298 
- 14.9228959779996928224 \, \sigma
- 359.335227793857981261 \, \sigma^{2}
+ e
}{
-0.529825425521888303709 
- 12.5121554370922843799 \, \sigma
+ 2.17885878641690129879 \, \sigma^{2}
}
\end{align}

where, $e=62.5743695601682449531\, \sigma^{3}$

\begin{align}
g(\sigma) = 
\frac{
-1.854927935283357414564
- 12.5788943758970529838 \, \sigma
- 195.424569686450992070 \, \sigma^{2}
+ h
}{
-0.306600502915980128356
- 4.67706630509962461331 \, \sigma
+ 0.58426006455412756666 \, \sigma^{2}
}
\end{align}

where, $h=24.4124444015615019915 \, \sigma^{3}$\\

\begin{figure}[ht]
\begin{center}
\includegraphics[width=.42\textwidth]{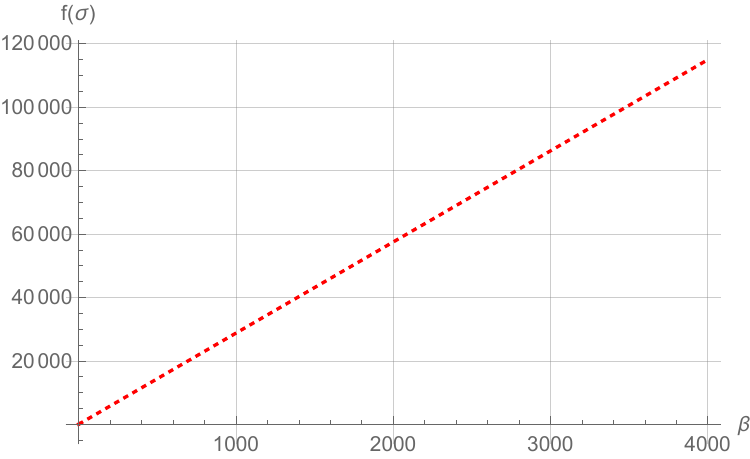}\vspace{3pt}
\includegraphics[width=.42\textwidth]{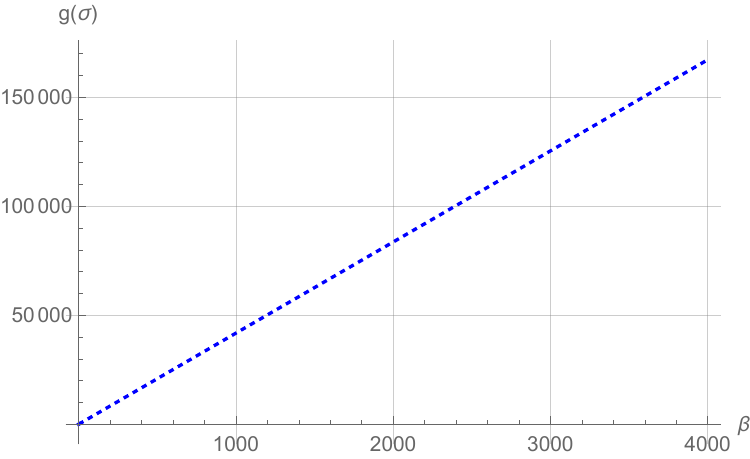}
\caption{$f(\sigma)$ and $g(\sigma)$ curves in the iso-$e$-charge processes for charged rotating black hole.}
\label{fig3}
\end{center}
\end{figure}

In Fig.\ref{fig3} , the behaviours of $f(\sigma)$ and $g(\sigma)$ are
depicted. Exact numerical results are represented by dots,

Now substituting the critical parameters to equation.\ref{eq14} and putting $\sigma=1$ we derive the critical parameters as:
\begin{equation}
S_c=0.68436~ C,~~~~\tilde{Q}_c= 0.023723 ~b ~C,~~~~~J= 0.023723~  C~~~~~T_c=\frac{0.269278}{l}
\end{equation}

It is noteworthy to mention that with appropriate adjustment of the parameters namely $\sigma=0$ we can deduce from here the critical parameters for the charged static black hole computed analytically in the previous case.\\
With,
\begin{equation}
t=\frac{T}{T_C},~~~~s=\frac{S}{S_C},~~~~~q=\frac{\tilde{Q}}{\sqrt{b}~\tilde{Q}_C}
\end{equation}
The relative temperature t and the rescaled Helmholtz free energy f for $b=2$ can therefore be written as follows:

\begin{align}
t = -\frac{0.0645782 \left(j^2 + 0.000562781 \, q^4 - 0.0474536 \, q^2 s^2 + 
    s^2 (-21.08 - 18.3681 \, s - 3.00097 \, s^2)\right) C^8}
{\sqrt{C^2} \, (s \, C^2)^{3/2} \, \sqrt{C^4} \, 
 \sqrt{0.0697989 \, j^2 (\pi + 0.68436 \, s) + 
        \left(0.0111088 \, q^2 + (2.14998 + 0.468349 \, s) s\right)^2 \, C^4}}
\end{align}

\begin{align}
f = \frac{9.38224 \left(0.0000201 \, q^4 + j^2 (0.0357 + 0.005185 \, s) + 
   q^2 (0.005185 + 0.000565 \, s) s + 0.2509 \, s^2 - 0.0119 \, s^4 \right) 
   C^4 \sqrt{C^2}}
{\sqrt{s C^2} \, \sqrt{C^4} \, 
 \sqrt{\left(0.0697989 \, j^2 (\pi + 0.68436 \, s) + 
        \left(0.0111088 \, q^2 + (2.14998 + 0.468349 \, s) s \right)^2 \right) C^4}}
\end{align}

It is important to mention that for $b > 1$ we enter in the realm of f(R) modified gravity away from the Einstein's theory.

\begin{figure}[ht]
\begin{center}
\includegraphics[width=.52\textwidth]{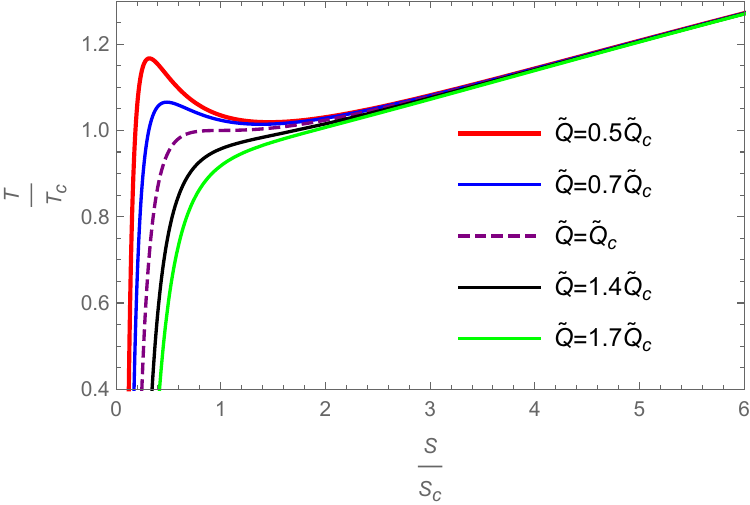}\vspace{3pt}
\includegraphics[width=.48\textwidth]{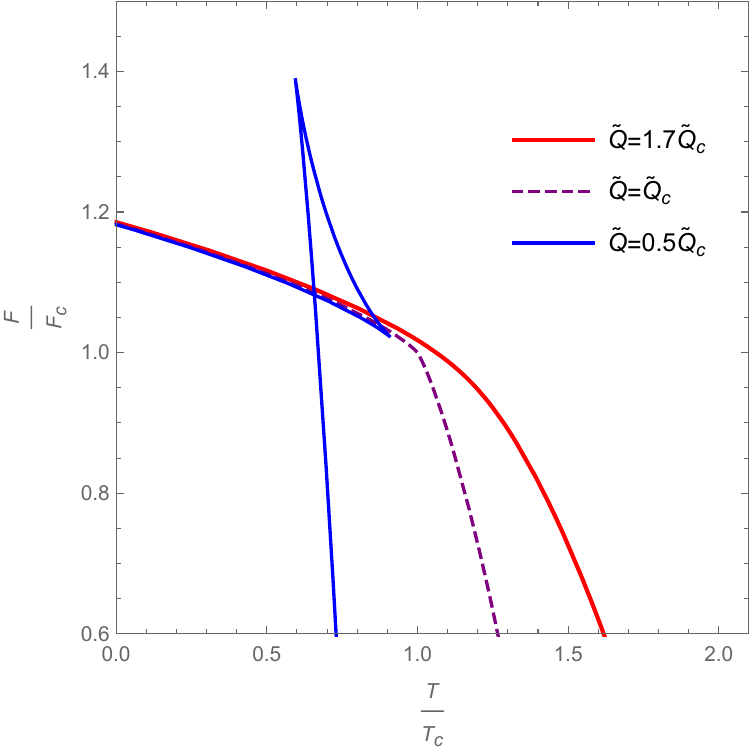}
\caption{$T-S$ and $F-T$ curves for charged rotating black hole in the iso-$e$-charge processes for $C=30$ and $b=2$.}\label{fig4}
\end{center}
\end{figure}

\begin{figure}[ht]
\begin{center}
\includegraphics[width=.48\textwidth]{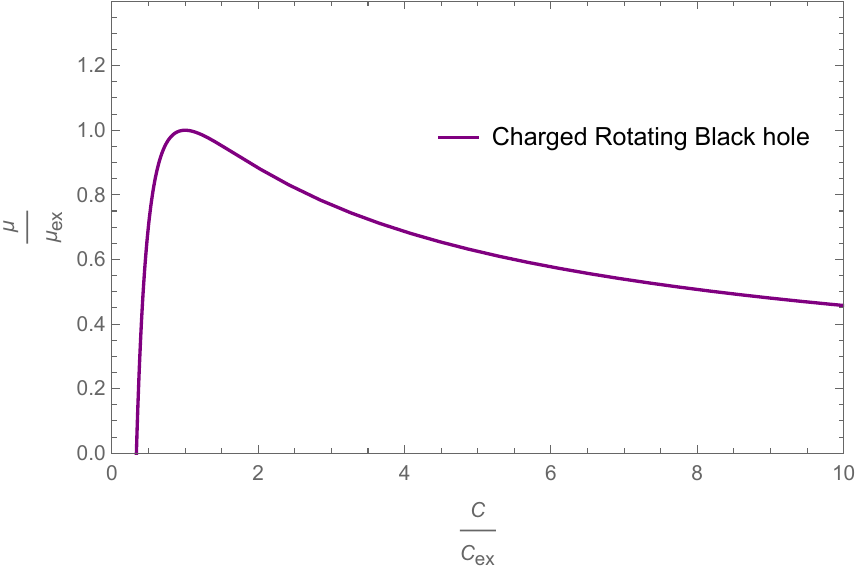}
\caption{$\mu-C$ curves for the charged rotating in the iso-$e$-charge processes for $C=30$ and $b=2$.}\label{fig5}
\end{center}
\end{figure}
 FIG. \ref{fig4} here illustrates both the $T-S$ and $F-T$ curves for the iso-e-charge process for the charged rotating black hole in f(R) gravity. The plots quite distinctly show that below the critical value of $\tilde{Q}_C$ both the $T-S$ and $F-T$ curves show non-trivial behaviour where, the $T-S$ curve exhibits non-monotonic behaviour below $\tilde{Q}_C$ whereas the $F-T$ curve displays a swallowtail behaviour below the critical value of $\tilde{Q}_C$. Such behaviour is an indication for the possible Van der Waals-like first-order phase equilibrium in the iso-e-charge processes when $0 < \tilde{Q} < \tilde{Q}_C$. However at the critical point $\tilde{Q} = \tilde{Q}_C$ we see that the second order phase transition is observed as represented by the dashed solid curve in both the figures.\\

We then analyze the $\mu-C$ process in detail for the iso-e-charge process. We numerically compute the extrememum value of $\mu$ i.e. $\mu_{ex}$  using eqtn.(\ref{eq9}), keeping $\tilde{Q}$,$J$,$b$, $C$ and $S$ fixed. We then draw a plot between $\frac{\mu}{\mu_{ex}}$ versus $\frac{C}{C_{ex}}$ and the corresponding plot is shown as  Fig.\ref{fig5}.

\section{Thermodynamic geometry of charged static and charged rotating black hole in $f(R)$ gravity}
\label{sec:GTD}

In geometrothermodynamics (GTD), the thermodynamic behavior of a system is described through differential geometry, with the aim of capturing phase transitions as geometric features \cite{Quevedo2007,GarciaPelaez:2014}. The framework is built on a thermodynamic phase space $\mathcal{T}$ of dimension $2n+1$, endowed with coordinates $Z^A=\{\Phi,E^a,I^a\}$, where $\Phi$ is a thermodynamic potential, $E^a$ are extensive variables, and $I^a$ their conjugate intensive variables $(a=1,\dots,n)$. This space is equipped with the contact 1-form 
\begin{equation}
\Theta = d\Phi - I_a\, dE^a ,
\end{equation}
which encodes the first law of thermodynamics. A Legendre invariant metric $G$ is then introduced on $\mathcal{T}$, for example, 
\begin{equation}
G = (d\Phi - I_a dE^a)^2 + (\delta_{ab} E^a I^b)\,\eta_{cd}\, dE^c dI^d ,
\end{equation}
with $\eta_{cd}=\text{diag}(-1,1,\dots,1)$, ensuring invariance under Legendre transformations \cite{Quevedo:2010curvature,Quevedo:2017homogeneity}. The space of equilibrium states $\mathcal{E}$ is defined via the smooth embedding $\varphi:\mathcal{E}\to\mathcal{T}$ given by 
\begin{equation}
\varphi: \{E^a\} \mapsto \{\Phi(E^a),E^a,I^a=\partial \Phi/\partial E^a\},
\end{equation}
such that $\varphi^*(\Theta)=0$, which corresponds to the Gibbs relation. The induced metric on $\mathcal{E}$, 
\begin{equation}
g = \varphi^*(G),
\end{equation}
then encodes the thermodynamic interaction: flat geometry ($R[g]=0$) corresponds to no thermodynamic interaction (ideal gas), while curvature singularities signal critical points and phase transitions \cite{QGas:2023,GTD:BH2021,Luongo:2023,Beissen:2023}. Thus, GTD provides a Legendre-invariant geometric criterion to study thermodynamic stability and critical phenomena.

\subsection{Charged static black hole}

\begin{figure}[h]	
	\centering
	\begin{subfigure}{0.37\textwidth}
		\includegraphics[width=\linewidth]{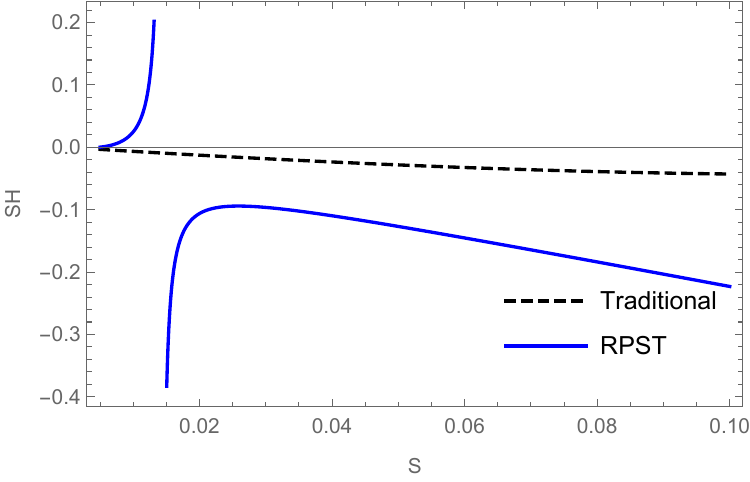}
		\caption{For $0<S<0.1$}
		\label{6a}
		\end{subfigure}
		\hspace{0.5cm}
		\begin{subfigure}{0.40\textwidth}
		\includegraphics[width=\linewidth]{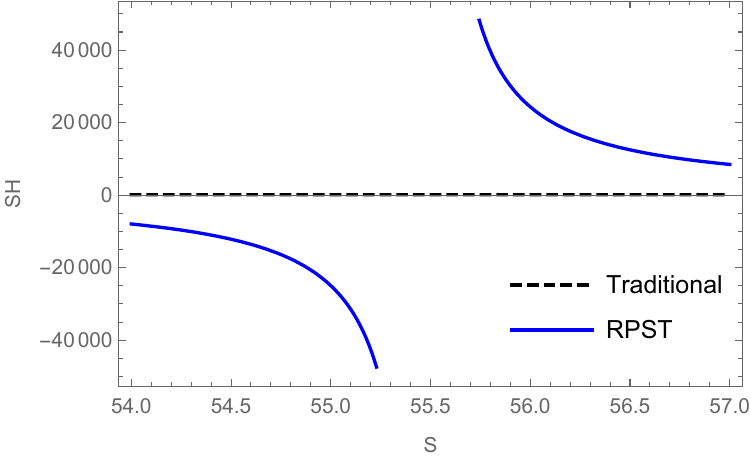}
		\caption{For $S > 0.1$}
		\label{6b}
		\end{subfigure}
		\caption{The plot for specific heat capacity for charged static BH in f(R) gravity in the iso-e-charge process }
	\label{6}
 \end{figure}

Following the specific heat capacity for the charged static black hole in f(R) gravity as has been given in equation.\ref{eq8} we can draw  plots for the specific heat capacity versus entropy for particular values of $C,\tilde{Q},b$ namely: $C=53$ and $\tilde{Q}=0.4$ and $b=2$. We find that the specific heat capacity for charged static black hole has discontinuities at $S=0.0142$ and $S=55.49$ respectively as can be seen in Fig.\ref{6}

Following the metric mentioned in equation.\ref{eq2} we can write the type II GTD metric for the charged static black hole in f(R) gravity as:

\begin{equation}
g  =S \left(\frac{\partial M}{\partial S}\right)\left(- \frac{\partial^2 M}{\partial S^2} dS^2 + \frac{\partial^2 M}{\partial C^2} dC^2 + \frac{\partial^2 M}{\partial \tilde{Q}^2} d\tilde{Q}^2 \right) 
\end{equation}

\begin{figure}[h]	
	\centering
	\begin{subfigure}{0.40\textwidth}
		\includegraphics[width=\linewidth]{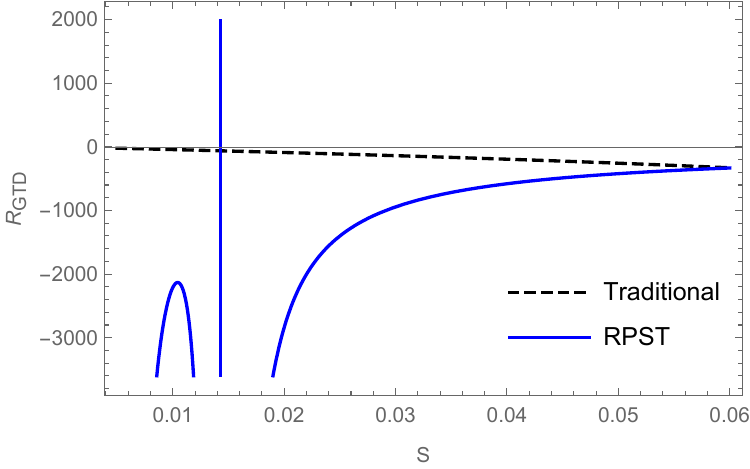}
		\caption{For $0<S<0.1$}
		\label{7a}
		\end{subfigure}
		\hspace{0.5cm}
		\begin{subfigure}{0.38\textwidth}
		\includegraphics[width=\linewidth]{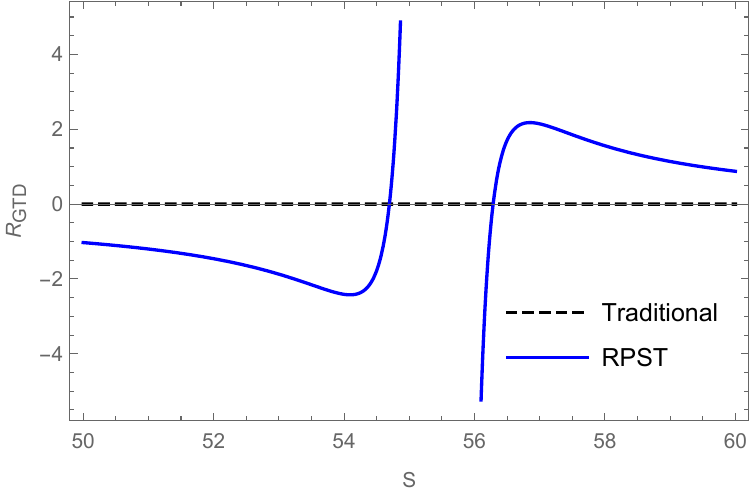}
		\caption{For $S > 0.1$}
		\label{7b}
		\end{subfigure}
		\caption{The plot for GTD scalar for charged static BH in f(R) gravity in the iso-e-charge process }
	\label{7}
 \end{figure}

And we compute the metric with the help from the expression for mass for the charged static black hole as obtained from the equation.\ref{eq6}and from which the GTD scalar can be computed which we avoid writing here due to its considerable length. When we draw this GTD scalar plot against entropy for the charged static case for fixed $C,\tilde{Q},b$ namely: $C=53$ and $\tilde{Q}=0.4$ and $b=2$ as can be seen from Fig.\ref{7}. We find that the divergences that occur in the GTD scalar curves for the charged static black hole in f(R) gravity matches exactly with those of the discontinuities in the specific heat capacity curves for the same black hole as can be seen from the previous plots.

\subsection{Charged rotating black hole}

\begin{figure}[h]	
	\centering
	\begin{subfigure}{0.37\textwidth}
		\includegraphics[width=\linewidth]{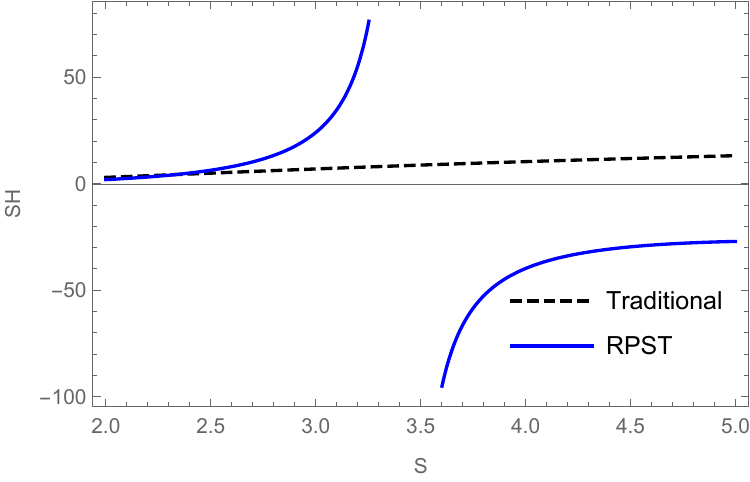}
		\caption{For $2 < S < 5$}
		\label{8a}
		\end{subfigure}
		\hspace{0.5cm}
		\begin{subfigure}{0.40\textwidth}
		\includegraphics[width=\linewidth]{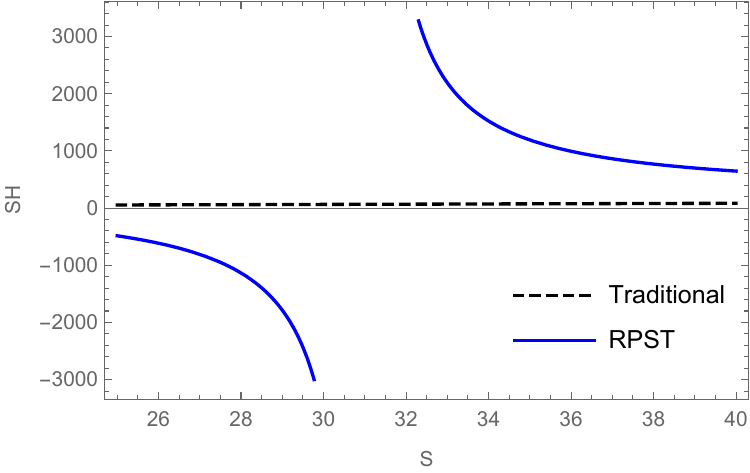}
		\caption{For $S > 5$}
		\label{8b}
		\end{subfigure}
		\caption{The plot for specific heat capacity for charged rotating BH in f(R) gravity in the iso-e-charge process }
	\label{8}
 \end{figure}

Following the specific heat capacity for the charged rotating black hole in f(R) gravity as has been given in equation.\ref{eq18} we can draw  plots for the specific heat capacity versus entropy for particular values of $C,\tilde{Q},J,b$ namely: $C=30$, $\tilde{Q}=0.2, J=0.2$ and $b=2$. We find that the specific heat capacity for charged rotating black hole has discontinuities at $S=3.42$ and $S=31.00$ respectively as can be seen in Fig.\ref{8}

Following the metric mentioned in equation.\ref{eq2} we can write the type II GTD metric for the charged rotating black hole in f(R) gravity as:

\begin{equation}
g  =S \left(\frac{\partial M}{\partial S}\right)\left(- \frac{\partial^2 M}{\partial S^2} dS^2 + \frac{\partial^2 M}{\partial C^2} dC^2 + \frac{\partial^2 M}{\partial J^2} dJ^2 + \frac{\partial^2 M}{\partial \tilde{Q}^2} d\tilde{Q}^2 \right) 
\end{equation}

\begin{figure}[h]	
	\centering
	\begin{subfigure}{0.40\textwidth}
		\includegraphics[width=\linewidth]{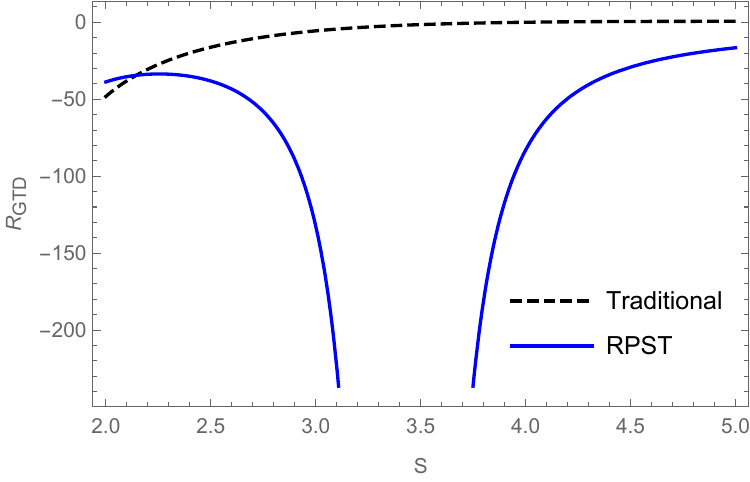}
		\caption{For $2< S <5$}
		\label{9a}
		\end{subfigure}
		\hspace{0.5cm}
		\begin{subfigure}{0.38\textwidth}
		\includegraphics[width=\linewidth]{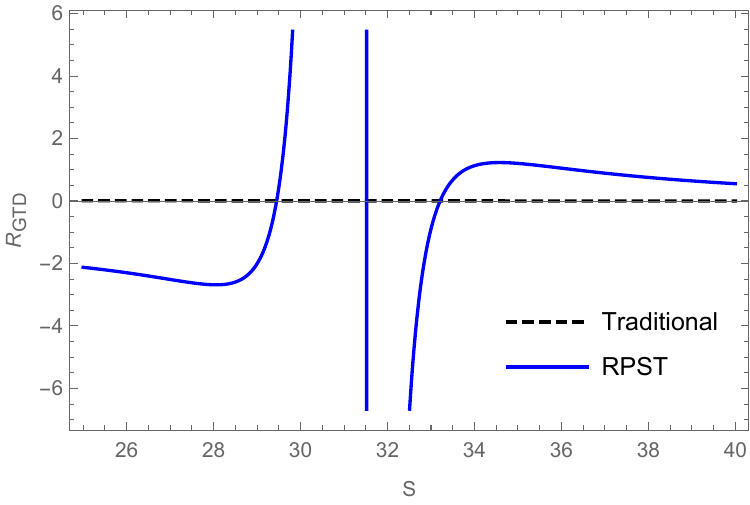}
		\caption{For $S > 5$}
		\label{9b}
		\end{subfigure}
		\caption{The plot for GTD scalar for charged rotating BH in f(R) gravity in the iso-e-charge process }
	\label{9}
 \end{figure}

And we compute the metric with the help from the expression for mass for the charged rotating black hole as obtained from the equation.\ref{eq13} and from which the GTD scalar can be computed which we avoid writing here due to its considerable length. When we draw this GTD scalar plot against entropy for the charged rotating case for fixed $C,\tilde{Q},J,b$ namely: $C=30$ and $\tilde{Q}=0.2, J=0.2$ and $b=2$ as can be seen from Fig.\ref{9}. We find that the divergences that occur in the GTD scalar curves for the charged rotating black hole in f(R) gravity matches exactly with those of the discontinuities in the specific heat capacity curves for the same black hole as can be seen from the previous plots.

\newpage
\section{Conclusion}
\label{sec:conclusion}

In this work, we have investigated the restricted phase space thermodynamics (RPST) of charged static and charged rotating black holes in $f(R)$ gravity. Motivated by the fact that $f(R)$ theories represent one of the simplest and most extensively studied modifications of general relativity, we sought to investigate whether the structural features of RPST, previously established primarily in the context of Einstein's gravity, do hold in such extended frameworks. Our results provide strong evidence that RPST does retain its predictive power and consistency even in the presence of higher-order curvature corrections.\\

For the charged static black hole, we derived explicit expressions for the thermodynamic quantities and analysed their scaling properties, confirming the first-order homogeneity of the mass in the restricted phase space. The $T-S$ and $F-T$ diagrams revealed the characteristic signals of phase transitions: the $T-S$ curves displayed non-monotonic behaviour below the critical charge, while the $F-T$ plots exhibited swallow-tail structures, both of which are distinct markers of first-order phase transitions akin to Van-der-Waals fluids. At the critical point, these features change into a second-order phase transition, thereby reproducing the universal phenomenology of black hole criticality.\\
 
In the case of charged rotating black holes, the situation was richer due to the interplay between electric charge and angular momentum. The analytical determination of critical points proved intractable, but through a careful numerical analysis we identified scaling relations and extracted critical quantities in terms of suitable dimensionless parameters. And just like the previous case, the hallmark features of first and second-order phase transitions re-emerged in the thermodynamic diagrams, thereby demonstrating that RPST is equally capable of capturing the criticality of more complex black hole space-times in $f(R)$ gravity.\\

To further validate these thermodynamic results, we employed the geometric formalism of geometrothermodynamics (GTD). By computing the Legendre-invariant GTD scalar curvature for both static and rotating cases, we were able to see a one-to-one correspondence between curvature singularities and divergences in the specific heat. This geometric-thermodynamic concordance reinforces the interpretation of GTD as a powerful and universal tool for phase transitions in black holes, particularly in modified gravity settings where standard methods may become excrutiating.\\

Overall, our investigation demonstrates that RPST provides a conceptually clean and mathematically consistent framework for analysing the thermodynamics of black holes in $f(R)$ gravity. The fact that it ably reproduces phase structures and critical phenomena analogous to those found in Einstein's gravity strongly suggests that RPST captures universal aspects of black hole thermodynamics, independent of the underlying gravitational theory. Moreover, the complementary role of GTD highlights the value of combining thermodynamic and geometric approaches in order to obtain a more complete understanding of critical behaviour.\\ 

As future aspects, it would be natural to extend this analysis to other modified gravity theories such as Gauss–Bonnet, Lovelock, or $f(T)$ gravity, as well as to higher-dimensional spacetimes where richer critical structures may tend to emerge. Furthermore, the holographic interpretation of the central charge in RPST opens up the possibility of exploring new connections with that of the dual conformal field theories, potentially shedding light on the microscopic origins of black hole entropy in modified gravity. We hope that the present work serves as a stepping stone in this direction, emphasizing the universality of RPST and its capacity to unify diverse aspects of black hole thermodynamics and geometry beyond Einstein’s theory of gravity.

\section{Acknowledgments}
	The authors would like to thank Pabitra Tripathy for the help he offered during the course of this work.

	\end{document}